\documentclass[preprint,twocolumn,10pt,a4paper,byrevtex,aip,superscriptaddress,showpacs]{revtex4-1}
\usepackage{natbib}
\usepackage{graphicx}
\usepackage{amsfonts}
\usepackage{amsmath}
\usepackage{amssymb}
\usepackage{color}

\begin{document}

\title{Band structure of topological insulators from noise measurements in tunnel junctions}

\author{Juan Pedro Cascales*}
\email{(*)juanpedro.cascales@uam.es}
\affiliation{Dpto. Fisica Materia Condensada C3, Instituto Nicolas Cabrera (INC), Condensed Matter Physics Institute (IFIMAC), Universidad Autonoma de Madrid, Madrid 28049, Spain}

\author{Isidoro Mart\'inez}
\affiliation{Dpto. Fisica Materia Condensada C3, Instituto Nicolas Cabrera (INC), Condensed Matter Physics Institute (IFIMAC), Universidad Autonoma de Madrid, Madrid 28049, Spain}

\author{Ferhat Katmis}
\affiliation{Department of Physics, Massachusetts Institute of Technology, Cambridge, Massachusetts 02139, USA}
\affiliation{Francis Bitter Magnet Laboratory, Massachusetts Institute of Technology, Cambridge, Massachusetts 02139, USA}

\author{Cui-Zu Chang}
\affiliation{Department of Physics, Massachusetts Institute of Technology, Cambridge, Massachusetts 02139, USA}

\author{Rub\'en Guerrero}
\affiliation{Instituto Madrile\~no de Estudios Avanzados en Nanociencia (IMDEA-Nanociencia), Cantoblanco, 28049 Madrid, Spain}

\author{Jagadeesh S. Moodera}
\affiliation{Department of Physics, Massachusetts Institute of Technology, Cambridge, Massachusetts 02139, USA}
\affiliation{Francis Bitter Magnet Laboratory, Massachusetts Institute of Technology, Cambridge, Massachusetts 02139, USA}

\author{Farkhad G. Aliev\textdaggerdbl}
\email{(\textdaggerdbl) farkhad.aliev@uam.es}
\affiliation{Dpto. Fisica Materia Condensada C3, Instituto Nicolas Cabrera (INC), Condensed Matter Physics Institute (IFIMAC), Universidad Autonoma de Madrid, Madrid 28049, Spain}

\begin{abstract}

The unique properties of spin-polarized surface or edge states in topological insulators (TIs)
make these quantum coherent systems interesting from the point of view of both fundamental physics and their
implementation in low power spintronic devices. Here we present such a study in TIs, through tunneling and noise 
spectroscopy utilizing TI/Al$_2$O$_3$/Co tunnel 
junctions with bottom TI electrodes of either Bi$_2$Te$_3$ or Bi$_2$Se$_3$. We demonstrate that
features related to the band structure of the TI materials show up in the
tunneling conductance and even more clearly through low frequency noise measurements. The bias dependence
of 1/f noise reveals peaks at specific energies corresponding to band structure features of the TI.
TI tunnel junctions could thus simplify the study of the properties of such quantum coherent systems, 
that can further lead to the manipulation of their spin-polarized properties for technological purposes. 

\end{abstract}

\maketitle

A topological insulator (TI) is a material which is insulating in the bulk but presents spin-dependent 
conducting edge or surface states which are protected by time-reversal symmetry \cite{Kane2005,Fu2007,Konig2007}.
A 2D or 3D TI presents edge or surface states respectively, which are spin-polarized in-plane, and locked at 
right angles to the carrier momentum, so that electrons with spin-up/down propagate in opposite directions. 
The edge or surface states of a TI consist of an odd number of massless Dirac cones. These properties along
with their high mobility make TI materials interesting for next generation, low dissipation, spintronic applications \cite{Chang2014high,TI_TMR} 
in which the electron spins are manipulated even without any magnetic fields. The experimental surge 
regarding these materials occurred with the prediction of Bi-based TIs \cite{arpes_calculations} and their 
posterior experimental realization \cite{arpes_bise}. Bi$_2$Se$_3$ and Bi$_2$Te$_3$, in particular, became
the prototypical TI materials that were studied most heavily.  

To date, the experimental verification of the band structure of TI materials has been 
predominantly carried out by angle-resolved photoemission spectroscopy (ARPES), which yields 
energy-momentum graphs of band dispersion for probing depths of under a few nm \cite{arpes_bisb,arpes_bise}. 
Also, the use of spin-ARPES has allowed the determination of the spin dependence of the topological surface states \cite{spin_arpes}. 
On the other hand, the use of scanning tunneling 
microscope (STM) allows obtaining information regarding the local density of states (DOS) and 
the topography of surfaces. By the study of quasiparticle scattering with STM, bands can be mapped very 
close to the Fermi surface, with a considerably lower energy range and at a smaller scale than with ARPES 
\cite{stm_ti,STM_BiTe_2,Cheng2010,Hanaguri2010}. Although immensely useful, these techniques are usually cumbersome and the conditions
of study are far from a practical application of TIs. A versatile and relatively simple technique to determine the DOS
of TI materials could be the study of electronic transport and noise 
through planar tunneling devices. So far, individual or heterostructure devices with TI layers have mainly 
dealt with lateral electron transport \cite{Checkelsky2011,Kong2011,Cho2011,Li2014,Assaf2014,Assaf2015,Liao2014,Jiang2014,
Wei2013,Assaf2013}. For certain spintronic applications, perpendicular transport may be required 
for which a direct or indirect contact between the TI
and other ferromagnetic layers is needed. Determining the band structure of TI films buried in multilayer structures
as well as confirming the robustness 
of their electronic properties remains an unresolved, central issue for the possible technological 
application of TIs in spintronic devices. Characterizing the bias dependence of low 
frequency noise in tunnel junctions can be a useful tool to gather information on the band 
structure of buried interfaces \cite{Aliev2014}, proving to be more sensitive to the opening and closing of
transmission channels than transport or inelastic tunneling spectroscopy (IETS) measurements.

\begin{center}
\begin{figure}[h]
    \begin{center}
    \includegraphics[width=\columnwidth]{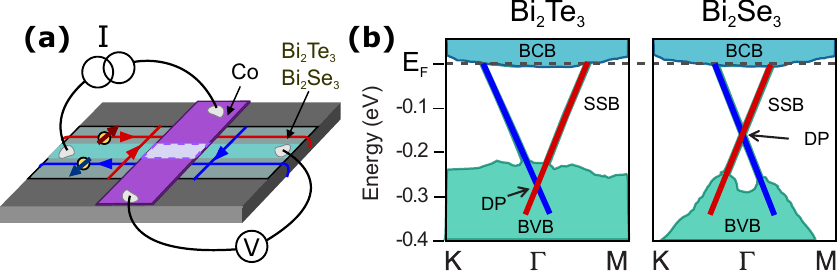}
    \caption{(a) Diagram of the sample structure. (b) Diagram of the band structure 
    of Bi$_2$Te$_3$ and Bi$_2$Se$_3$, adapted from Refs. \cite{cuizubite,cuizubise}} \label{fig:fig1}
    \end{center}
\end{figure}
\end{center}

In this letter we report on the investigation of the band structure of the TI electrodes 
in perpendicular tunnel junctions by both electron transport and low frequency noise spectroscopy. 
Having a perpendicular tunneling transport 
allows probing the DOS of both the surface states and the bulk bands of the materials. 
The bottom TI electrodes of the samples are Bi$_2$Te$_3$ and Bi$_2$Se$_3$ with thicknesses 
of 10 and 20 quintuple layers (QL), well into the 3D TI range \cite{2d3dTIs,cuizu2d3d}. 
A schematic diagram of the sample structure is shown in Fig. \ref{fig:fig1}(a). In Bi$_2$Te$_3$, the Fermi 
energy lies within the conduction band and the Dirac point of the surface states is in the valence band. 
Bi$_2$Te$_3$ presents a similar band structure but the Dirac point is located between the valence and 
conduction bands. A schematic diagram of the band structure, adapted from ARPES results of Bi$_2$Te$_3$ 
\cite{cuizubite} and Bi$_2$Se$_3$ \cite{cuizubise} shown in Fig. \ref{fig:fig1}(b), where BCB stands for 
bulk conductance band, BVB for bulk valence band, SSB for surface state band and DP for Dirac point.

\begin{figure}[h]
    \begin{center}
    \includegraphics[width=\columnwidth]{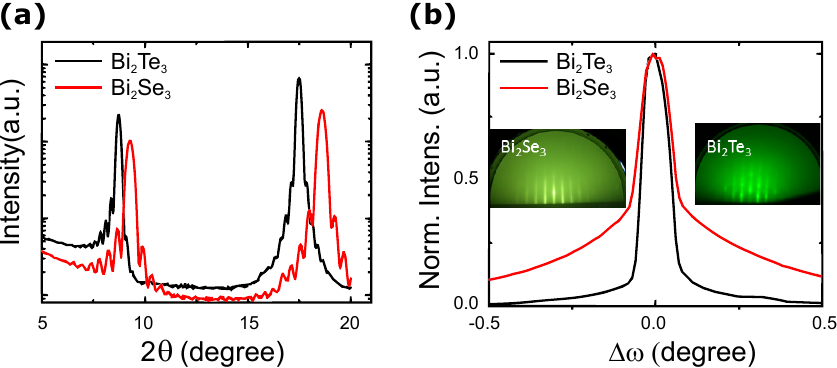}
    \caption{Cu-K$_{\alpha}$ radiation and 15 keV electron beam are used for XRD and RHEED 
    measurements, respectively. (a) Typical XRD pattern for symmetric Bragg reflections of 
    Bi$_2$Se$_3$ (in red) and Bi$_2$Te$_3$ (in black). (000l)-type reflections are visible 
    with pronounced thickness fringes at the vicinity of the layer reflections which shows an 
    (000l)-type layer orientation. The fringes along the out-of-plane direction are due to the layer thickness 
    and the quality of the film. (b) In-plane line-cut 
    for symmetric (0006) reflection for both type of TI films. The rocking curve for (0006) 
    Bragg peak is taken along the $\left\langle \bar{1}010 \right\rangle$ 
    crystallopraphic direction of the substrate. The insets show the RHEED images 
    for both TI films.} \label{fig:fig2}
    \end{center}
\end{figure}

The Bi$_2$Te$_3$ samples on Si(111) substrates have a layer structure of: Bi$_2$Te$_3$(10 QL) / Al$_2$O$_3$(5.8 nm)/ 
Co(100 nm)/ AlO$_x$(3.8 nm). In order to improve the TI/barrier interface, we then employed (0001)-oriented, epi-ready, 
commercial Al$_2$O$_3$ (sapphire) substrates over which a thicker, 20 QL film of Bi$_2$Se$_3$ 
was grown. The layer sequence for the Bi$_2$Se$_3$ junctions on sapphire substrates is: Bi$_2$Se$_3$ (20 QL)
/ Al$_2$O$_3$(5.7 nm) / Co(100 nm)/ AlO$_x$(3.8nm). The TI thin film growth was carried out by a molecular beam epitaxy (MBE) apparatus under 
ultra-high vacuum environment ($10^{-9}-10^{-10}$ Torr). High purity (5N) elemental Bi, Te, 
and Se were evaporated at growth rates between $0.5-1$ nm/min, after which the film was annealed for an hour at 600 $^{\circ}$C,
and at 800 $^{\circ}$C for 30 min under $10^{-9}$ Torr. 
In-situ reflection high-energy electron diffraction (RHEED) monitorization (Fig.\ref{fig:fig2}(b) inset), indicates a good two-dimensional 
growth. The X-ray diffraction (XRD) pattern ($\omega-2\theta$) of the films along 
the growth direction for 20 QL Bi$_2$Se$_3$ (in red) and Bi$_2$Te$_3$ (in black) are shown in Fig. \ref{fig:fig2}(a). 
Laue oscillations around the Bragg peaks ((0003) and (0006)) are a clear 
indication of the film's structural coherence along the growth direction all the way to the top surface. 
Fig.\ref{fig:fig2}(b) shows a rocking curve ($\omega$-scan) on each Bragg peak and one of the (0006)-peaks, where
Gaussian broadening is caused by finite size effects and an additional Lorentzian lineshape appears due to 
defects, such as vacancies, dislocations, etc. By fitting a pseudo-Voigt function to the curves, 
we can conclude that the defect density within Bi$_2$Se$_3$ is larger than for Bi$_2$Te$_3$.

\begin{center}
\begin{figure}[!h]
    \begin{center}
    \includegraphics[width=\columnwidth]{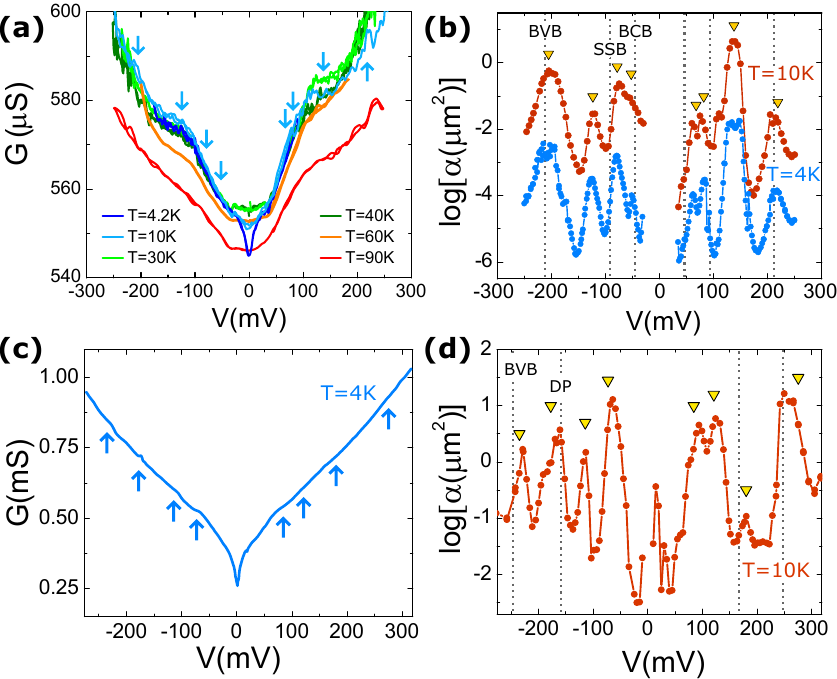}
    \caption{(a) Conductance measurements at different temperatures for a Bi$_2$Te$_3$/AlO$_x$/Co tunnel junction. 
    (b) 1/f noise $vs.$ bias in the Bi$_2$Te$_3$ junction for two temperatures (curves offset for convenience), correlated with 
    inflection points of the conductance (triangles) and Bi$_2$Te$_3$ band features (dashed lines)\cite{stm_ti}. 
    (c) Conductance at T=4 K for a Bi$_2$Se$_3$/AlO$_x$/Co tunnel junction and (d) 1/f noise at T=10 K, 
    compared to conductance inflection points (triangles) and Bi$_2$Se$_3$ band features (dashed lines)\cite{cuizubise}.} \label{fig:fig3}
    \end{center}
\end{figure}
\end{center}

The transport and low frequency noise setup was described in previous articles \cite{Guerrero2006,Guerrero2007}. 
Electrons tunnel from the Co (TI) to the TI (Co) for positive (negative) bias.
Fig. \ref{fig:fig3} presents low temperature, transport measurements carried out on a 1.8 k$\Omega$ Bi$_2$Te$_3$ sample. 
A fit of the Brinkman's \cite{Brinkman} (not shown) to the sample's IV at T=0.3 K with yields an effective 
tunnel barrier thickness of around 1.5 nm. This could imply a rough TI/barrier interface, since the nominal 
barrier is 5.8 nm thick. As can be seen in Fig. \ref{fig:fig3}(a), the conductance presents a parabolic-like 
dependence (junction-like behavior) with changes in slope at energies which could be related to features 
in the DOS of the TI. The overall shape of the conductance did not change with the temperature, except for 
a zero-bias anomaly peak which accentuated with decreasing temperature. This could mean that the 
surface states are robust at least up to 90 K. The slope changes in the conductance could correspond, as was
discussed in Ref. \cite{stm_ti}, to leaving and entering the different bands of the materials: conductance 
band, surface state and valence band.

\begin{center}
\begin{figure}[h]
    \begin{center}
    \includegraphics[width=\columnwidth]{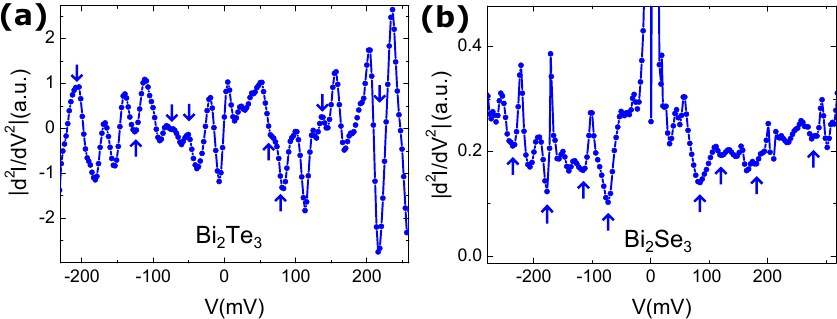}
    \caption{Derivative (in absolute value) of the conductance at T=4 K of (a) Bi$_2$Te$_3$ and (b) Bi$_2$Se$_3$ junctions. Arrows indicate the bias of maxima or minima close to peaks in 1/f.} \label{fig:fig4}
    \end{center}
\end{figure}
\end{center}

1/f noise is described by the power spectral density $S_V(f)=\frac{\alpha V^2}{A f^\beta}$\cite{Hooge} where
$V$ is the applied bias, $A$ the junction area and $\beta\sim 1$. The normalized 1/f noise or Hooge parameter 
$\alpha$ is extracted from the spectra by performing the linear fit 
$log\left(S_V(f)\right)=log\left(\frac{\alpha V^2}{A}\right)-\beta log(f)$.
As is shown in Fig. \ref{fig:fig3}(b), $\alpha$ presents several clear peaks at certain values of energy which compare favorably 
with inflection points in the conductance curves, shown as arrows in Fig. \ref{fig:fig3}(a) 	
and as triangles in (b). The energies are estimated from the IETS spectrum in Fig. \ref{fig:fig4}(a),
presenting features which could arise from phonons in the AlO$_x$\cite{Dragoset2011}, 
the TI \cite{Zhu2011} or coupled AlO$_x$-TI modes. The occurrence of inelastic tunneling processes with the addition of
sequential tunneling observed in shot noise measurements (not shown) could lead to a fraction of
the carriers being inelastically back-scattered against the flow of the current\cite{Buttiker1986}. This 
could explain why the peaks in 1/f noise are observed for both signs of the bias.
Furthermore, these energies can be seen to relate to the band structure of 
the TI obtained by ARPES and STM in Ref. \cite{stm_ti}, shown by dashed lines. The 1/f peak around 
$\pm 200$ mV is close to the value for the beginning of the valence band \cite{stm_ti} (BVB).  
The remaining peaks could be due to the opening TI surface state band (SSB) and the bulk conductance 
band (BCB). These features may originate from the opening or closing of transport channels which
influences the voltage fluctuations, as shown in Ref. \cite{Aliev2014}. The temperature dependence of the 
peaks around $-100$ mV was tracked (not shown) when the sample was allowed to warm up from LHe$^{4}$ 
temperatures. The band-related features gradually disappear when the temperature is increased. However, 
the position of both peaks, specially the one around -120 mV, does not shift in energy when the temperature
increases. This may provide an argument that these features are related to the surface states, which have 
been reported to be thermally stable\cite{Zhao2014}.

Similar results were obtained for the junctions with a 20 QL Bi$_2$Se$_3$ electrode. 
Fig. \ref{fig:fig3}(c)(d) present the results obtained for a 4.2 k$\Omega$ junction. The dependence 
of the conductance with the bias is shown in Fig. \ref{fig:fig3}(c), which is parabolic-like and
presents changes in slope, which can be estimated by differentiating the conductance curve (Fig. \ref{fig:fig4}(b)). 
We indicate the most relevant inflection points for 1/f results by arrows. The analysis of 1/f noise also presents peaks which appear at different 
energies (see Fig. \ref{fig:fig3}(d)). Inflection points in the conductance from Fig. \ref{fig:fig4}(b), 
which are close in energy to the peaks in 1/f are indicated by 
triangles and compared to the band features of the TI from Ref. \cite{arpes_bise}.  The peak located around 
$\pm 250$ could be related to the beginning of the valence band (BVB). 
The feature located around $\pm 160$ mV could be related to the Dirac point (DP). The remaining peaks in 
1/f noise could be due to the opening of the surface state band 
and the conductance band, as in the Bi$_2$Te$_3$ sample.

Even though evidence of transport through the TI surface states is found, we did not observe a 
magnetoresistive response with fields up to 3 kOe. Since the bulk DOS of TIs is not polarized in spin, the Zeeman 
splitting caused by the 3 kOe field might not be enough to observe any significant effect. Whereas the 
absence of the surface driven spin polarization in the TI response may indicate a loss of spin 
polarization through spin scattering at the interfaces or defects \cite{Moodera1999,MooderaRev2011,MooderaChapter}.
The ZBA seen in the conductance $vs$ bias may 
relate to loss of spin information due to Co atoms inside 
the Al$_2$O$_3$ barrier which act as scattering centers or to higher surface 
defect densities of the TI, as shown by XRD. 

In conclusion, we report the fabrication of tunnel junctions with a TI bottom electrode. 
We have demonstrated that features related to the band structure of these materials in contact with ferromagnetic layers 
through an Al$_2$O$_3$ barrier can be experimentally detected through electronic transport 
measurements, and even more clearly through low frequency noise measurements. We do not exclude the 
influence of inelastic scattering by barrier and interface related phonons\cite{Drewello2008}. One way to confirm our results could be to look 
for peaks in 1/f noise in Al/Al$_2$O$_3$/Bi junctions
as Bi shows band features in the conductance $vs$ bias. Obtaining reliable, 
TI tunnel junctions could simplify the study of their properties and pave the way for the manipulation of 
their spin-polarized properties for technological purposes. Future measurements will deal with
TI films in the 2D limit, with the aim of exploiting the spin polarized edge states for spintronic
applications. An interesting alternative would be to carry out noise measurements
in $pn$ junctions with a TI semiconductor\cite{pn_ti}. Also, crystalline barriers such as 
MgO could improve the interface, and allow the probing of the surface state spin texture.

F.A., and I.M acknowledge support by the Spanish MINECO (MAT2012-32743), and the Comunidad de Madrid through NANOFRONTMAG-CM (S2013/MIT-2850), J.P.C. acknowledges support from the Fundacion Seneca (Region de Murcia) posdoctoral fellowship (19791/PD/15), and J.S.M., C.Z.C and F.K. from grants NSF (DMR-1207469), ONR (N00014-13-1-0301), and the STC Center for Integrated Quantum Materials under NSF grant DMR-1231319.

\end{document}